\documentclass[journal=jpccck,manuscript=article,layout=twocolumn]{achemso}
\setkeys{acs}{keywords = true}
\setkeys{acs}{usetitle = true}
\usepackage[version=3]{mhchem} 
\usepackage{color}
\usepackage{threeparttable}
\usepackage[fontsize=10pt]{fontsize}

\usepackage{lineno}
\usepackage{multicol}
\usepackage{hyperref}
\usepackage{layouts}
\author{Zhen Zeng}
\affiliation{State Key Laboratory of Engines, Tianjin University, Tianjin, 300072, China.}
\author{Kai Sun}
\affiliation{State Key Laboratory of Engines, Tianjin University, Tianjin, 300072, China.}
\author{Rui Chen}
\affiliation{Department of Aeronautical and Automotive Engineering, Loughborough University, Loughborough LE11 3TU, United Kingdom.}
\author{Mengshan Suo}
\affiliation{State Key Laboratory of Engines, Tianjin University, Tianjin, 300072, China.}
\author{Zhizhao Che}
\email{chezhizhao@tju.edu.cn}
\affiliation{State Key Laboratory of Engines, Tianjin University, Tianjin, 300072, China.}
\author{Tianyou Wang}
\email{wangtianyou@tju.edu.cn}
\affiliation{State Key Laboratory of Engines, Tianjin University, Tianjin, 300072, China.}

\title{Variation of Critical Crystallization Pressure for the Formation of Square Ice in Graphene Nanocapillaries}


\begin{document}

\section{ABSTRACT}
\includegraphics[width=1.05\columnwidth]{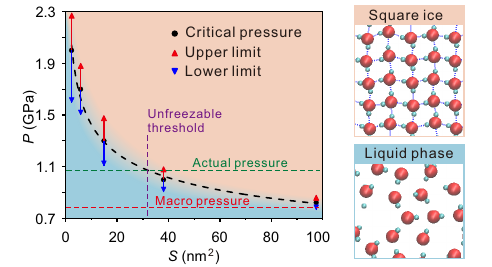}
Two-dimensional square ice in graphene nanocapillaries at room temperature is a fascinating phenomenon and has been confirmed experimentally. Instead of temperature for bulk ice, the high van der Waals pressure becomes an all-important factor to induce the formation of square ice and needs to be studied further. By all-atom molecular dynamics simulations of water confined between two parallel graphene sheets, which are changed in size (the length and the width of the graphene sheets) over a wide range, we find that the critical crystallization pressure for the formation of square ice in nanocapillary strongly depends on the size of the graphene sheet. The critical crystallization pressure slowly decreases as the graphene size increases, converging to approximately macroscopic crystallization pressure. The unfreezable threshold for graphene size is obtained by estimating the actual pressure and it is difficult to form the square ice spontaneously in practice when the graphene sheet is smaller than the threshold. Moreover, the critical crystallization pressure fluctuates when the graphene size is minuscule, and the range of oscillation narrows as the sheet size increases, converging to the macroscopic behavior of a single critical icing pressure for large sheets. The graphene size also affects the stability and crystallization time of square ice.

\section{Introduction}\label{sec:sec1}
The phase behavior of water is omnipresent and important in our daily life, and has many significant implications for physical and chemical processes \cite{Mishima1998SupercooledGlassyWater}. Up to now, more than 19 bulk ice phases have been confirmed by experiments and some others have been revealed from numerical simulation predictions \cite{Debenedetti2003SupercooledGlassyWater, Huang2016UnderNegativePressure, Poole1992MetastableWater, Sciortino1995CrystalStabilityLimits, Zhao2014TwoDimensionalAmorphousIce, Zheng1991HomogeneousNucleationLimit, Gasser2021Ice19}. However, confined water/ice is less considered but undoubtedly ubiquitous, which has many different characteristic and practical applications \cite{wang2019MonolayerIce, Compton2012MechanicalPropertiesofGraphene, he2019PhaseBehaviorConfinedWater, Neek2016ViscosityofNanoconfinedWater, kwac2017MultilayerWater}.

For water confined in nano space, many novel crystal and quasicrystal structures have been reported over the past two decades. By classical and ab-initio molecular simulations, some new ice phases have been revealed. Among them, low-dimensional ice phase is noteworthy for its unique constructions. For instance, plentiful quasi-one-dimensional crystal structures can be found when the water molecules are confined in carbon nanotubes, such as square \cite{Koga2000IceNanotube}, pentagonal \cite{Bai2003PentagonandHexagonIce, Koga2001CarbonNanotubes, Koga2000IceNanotube}, hexagonal \cite{Bai2003PentagonandHexagonIce, Koga2001CarbonNanotubes, Koga2000IceNanotube}, and heptagonal \cite{Bai2003PentagonandHexagonIce, Byl2006UnusualHydrogenBonding} single-walled ice nanotubes, or helical multi-walled ice nanotube with high density under higher hydrostatic pressure \cite{Bai2006MultiwalledIceHelixes}. Numerous two-dimensional (2D) monolayer and multilayer ice structures can be found in the confinement composed of two parallel sheets, such as octagonal \cite{Bai2010MonolayerClathrate}, tetragonal \cite{Bai2010MonolayerClathrate}, hexagonal \cite{Zhao2014FerroelectricHexagonal}, flat rhombic \cite{Zhao2014FerroelectricHexagonal}, puckered rhombic \cite{Bai2010MonolayerClathrate}, and square monolayer ices \cite{Algara2015SquareIce}. Multilayer ice can also present the same patterns \cite{Chen2017DoubleLayerIce, Corsetti2016HighDensityBilayer, Koga2005HydrophobicSurfaces, Koga2000AmorphousPhases} and some novel patterns \cite{Algara2015SquareIce, Bai2012PolymorphismandPolyamorphism, Zhu2015CompressionLimit}. Some new phases of crystalline ice have been confirmed experimentally. Systematic X-ray diffraction analysis has been carried out to identify different polygonal nanotube ice structures, and the results agree with those obtained from molecular dynamics (MD) simulations \cite{Maniwa2005PentagonaltoOctagonal}. With the help of high-resolution friction force microscope, experimental evidence has been provided to support the existence of 2D ice above the melting temperature of bulk ice \cite{Jinesh2008ExperimentalEvidence}. 2D square ice at room temperature has also been confirmed by transmission electron microscopy \cite{Algara2015SquareIce}.

The formation of the above novel crystalline structures can be attributed to the effect of confinement. In general, the matric suction can be interpreted using capillarity in relatively large pores (i.e., pore diameters greater than 10 nm, but still in nanoscale). In contrast, in minuscule pores (pore diameters less than 3 nm), the matric suction will be dominated by adsorption instead of capillarity \cite{Zhang2018UnfreezableThreshold}. The matric suction increases as the pore diameter decreases, indicating that a smaller pore can provide a greater confinement effect, and this effect works on the water molecules inside the pores and manifests itself in the form of ultrahigh pressure, which is important to induce the novel ice structures \cite{Algara2015SquareIce}. Among the plenty of crystal structures in nanoscale confinements, two-dimensional square ice (including nearly square ice) is noteworthy because this square structure is qualitatively different from the conventional tetrahedral structure and has been confirmed experimentally \cite{Algara2015SquareIce}. The square ice is almost insensitive to the hydrophobicity of the confining sheets, and its melting temperature is anomalously higher than that of the bulk ice \cite{Algara2015SquareIce}. In recent years, although there are some further studies by varying the thermodynamic variables \cite{Zhu2016AbStacked, Zhu2016TrilayerIces, Zhu2016BucklingFailureofSquareIce, Zhu2017SuperheatingofMonolayerIce}, the separation of graphene sheets \cite{Zhu2015CompressionLimit, Zhu2016TrilayerIces}, and the type of confining surfaces \cite{Qiu2015InhomogeneousNanoconfinement, Ruiz2018FlexibleConfiningSurfaces, goswami2020HighpressIce, bampoulis2016HydrophobicIce, kim2016IcyWaterClusters}, the variation of the critical crystallization pressure for the square ice formation still requires further investigation, which is all-important to the square ice. Instead of temperature, the high van der Waals pressure becomes an important factor to induce the square icing structure \cite{Algara2015SquareIce}, similar to the importance of temperature to the freezing process for bulk ice. Zhu et al.\ \cite{Zhu2015CompressionLimit} characterized the compression limit of the 2D monolayer water from a mechanics point of view, and obtained a compression-limit phase diagram. They focused on the effect of the height of the nanocapillary (the height of the capillary in this study, i.e., the separation between the two graphene sheets in $y$ direction), and only considered a very small range of the height of the nanocapillary (from 6.0 to 11.6 {\AA}) due to the stability limitation of the multilayer ice. It is worth noting that the multilayer ice has a strict limit on the number of layers, because it can keep stable only for a small range of sheet separations \cite{Zangi2004WaterConfinedReview}. Moreover, there are appropriate heights of the nanocapillaries for the formation of flat monolayer, bilayer, and trilayer square ice \cite{Zeng2022PartitionedSquareIce}. This determines that the diagram is discontinuous and can be divided into three parts, and the water molecules formed a variety of structures besides square ice (both crystalline-like and amorphous structures).

The existence of low-dimensional ice at room temperature has been proposed to explain the fast water permeation through nanocapillaries, including carbon nanotubes and graphene-based membranes. In some situations, the low-dimensional ice needs to be suppressed, such as the nanocapillaries in the catalytic layer of a proton exchange membrane fuel cell, in which the existence of low-dimensional ice will hinder the transfer of reactants and reduce the catalytic efficiency. It can be found that most of the previous studies on the crystallization pressure of square ice confined in graphene nanocapillaries have focused on the effects of the heights of nanocapillaries, and used minuscule graphene sheets with fixed sizes. It is often assumed that the critical crystallization pressure for the square ice formation is independent of the size of graphene sheets (i.e., the length and the width of the graphene sheets, corresponding to $D_z$ in $z$ direction and $D_x$ in $x$ direction in Figure \ref{fig:01}). However, in this study, we find that the critical crystallization pressure is not constant but strongly depends on the size of the graphene sheets, even at a fixed separation distance between the graphene sheets. We carry out all-atom molecular dynamics simulations of water in confinements composed of two parallel graphene sheets, which are changed in size over a wide range. Our results unveil the variation of critical crystallization pressure for the square ice formation, combined with the macroscopic perspective. Thus, the existence of low-dimensional ice can be facilitated or suppressed by the design of capillary length in practical applications, according to the interrelation between the length of capillary and critical crystallization pressure. This study provides physical insights about the variation mechanism of critical crystallization pressure for the square ice formation, and widespread applications in many fields such as nanomaterial \cite{Garcia2022interfacial}, nanofluidic \cite{negi2022Ice2D, Garcia2022interfacial}, and nanotribology \cite{Lin2019influence}.

\section{Computational methods}\label{sec:sec2}
\subsection{Molecular dynamics simulation}
\begin{figure}
  \centering
  \includegraphics[width=0.9\columnwidth]{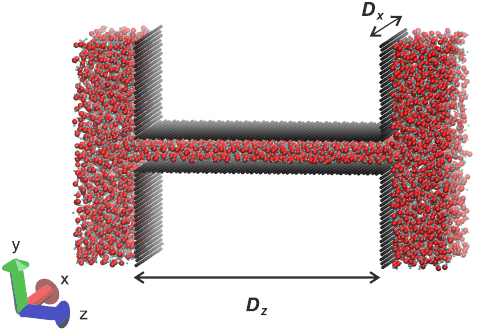}\\
  \caption{Configuration of MD simulation system. The red spheres indicate oxygen atoms, the cyan spheres indicate hydrogen atoms, and the black spheres indicate carbon atoms. The width ($D_x$ in $x$ direction) and the length ($D_z$ in $z$ direction) of the graphene sheets are varied.}\label{fig:01}
\end{figure}

Figure \ref{fig:01} is the configuration of our MD simulation system. Two water reservoirs are connected by a capillary, which is formed by two parallel graphene sheets. The length and the width of the graphene sheets ($D_z$ in $z$ direction and $D_x$ in $x$ direction) are varied. The height of the graphene capillary (i.e., the separation between the two graphene sheets in $y$ direction) is set to be 9.0 {\AA} to accommodate bilayer square ice. The coordinates of the carbon atoms in the graphene sheets are fixed in the simulation. Our MD simulations are carried out in the isothermal, isobaric ensemble, and the pressure and the temperature are controlled by the Nos\'{e}-Hoover barostat and thermostat. The three directions in the simulation are all set to periodic boundary conditions. The temperature is set constant ($T$ = 298 K), and the lateral pressure $P$ (the high pressure induced by encapsulating graphene sheets is modelled to a first approximation by applying a hydrostatic pressure $P$ in the direction parallel to the graphene layers) is increased stepwise at a fixed rate of 0.1 GPa/ns in the pressurization process. In our simulations, the $z$-length of the simulation box will change only according to the lateral pressure in $z$ direction, which will change the distance between the water molecules so that the pressure of the water molecules in $z$ direction will approaches the set lateral pressure. The box dimensions in $x$ and $y$ direction do not change in the simulations. So, there is no inappropriate contribution of vacuum to the pressure of water molecules. To facilitate the comparison of results, especially in the study of structural stability, we fixed the aspect ratio of the computing domain for graphene sheets. The sizes of graphene sheets in $z$ and $x$ directions are changed proportionally and chosen as 26.6 {\AA} $\times$ 21.9 {\AA}, 42.5 {\AA} $\times$ 35.0 {\AA}, 68.0 {\AA} $\times$ 56.0 {\AA}, 108.8 {\AA} $\times$ 89.6 {\AA}, respectively. As for the size effect of $D_x$, some additional calculations are performed. From the comparison of results with different $D_x$, the potential energy of water molecules is affected by the system-size effect in $x$ direction, both in the liquid and crystalline state. But the critical crystallization pressure is not significantly affected, as shown in Figure S1 in Supporting Information.

The extended simple point charge (SPC/E) model is chosen in our study, including the short-range Lennard-Jones potential and the long-range Coulomb potential \cite{Berendsen1987EffectivePairPotentials}. For the simulations of bilayer square ice structure, it has been proved that the SPC/E water model is accurate, by comparing with experimental results \cite{Algara2015SquareIce}. Other water model choices have been used for comparison in this study, as shown in Figure S2 in Supporting Information, and the simulation results show that different water models do not affect the main findings between graphene size and critical crystallization pressure. The 12-6 Lennard-Jones potential between the oxygen and the carbon atoms is chosen to model the water-carbon interaction. The energy and the distance parameters are $\epsilon$ = 0.114 kcal/mol and $\sigma$ = 0.328 nm \cite{Che2017SurfaceNanobubbles, Gordillo2000HydrogenBondStructure, Joly2011GiantLiquidSolidSlip, Werder2008WaterCarbonInteraction}. In our simulations, the cutoff distance for L-J potential is set to be 10.0 {\AA}. The particle-particle particle-mesh (PPPM) algorithm with an accuracy of $10^{-4}$ is used to compute the long-range interactions. A time step of 1.0 fs is used for the velocity-Verlet integrator. The simulations in this study are carried out by using LAMMPS \cite{Plimpton1995FastParallelAlgorithms}, and all the snapshots of simulation results are rendered in VMD \cite{Humphrey1996VMD}.

\subsection{Square icing parameters}
Square icing parameters, $MCV_1$ and $MCV_2$, are used in our study to quantitatively distinguish between the square ice structure and the liquid water. The parameters can analyze the uniformity of the molecule's distribution in the direction of the hydrogen-oxygen bonds and spatial position, and identify the different water phases when we take appropriate thresholds. See the Supporting Information and our previous work \cite{Zeng2022PartitionedSquareIce} for more detailed definitions and descriptions of the square icing parameters.

\section{Results and discussion}
\subsection{Square icing phenomenon for all graphene sizes}
\begin{figure*}
  \centering
  \includegraphics[width=1.95\columnwidth]{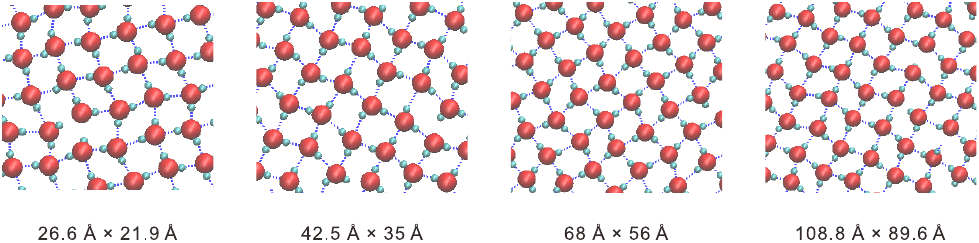}\\
  \caption{Top views of the final simulation snapshots for graphene capillaries of different sizes. The red spheres indicate oxygen atoms, the cyan spheres indicate hydrogen atoms, and the blue dashed lines indicate hydrogen bonds.}\label{fig:02}
\end{figure*}

 We carry out a series of all-atom molecular dynamics simulations of the pressurization process to capture the transition of water molecules from liquid to crystal. Temperature is set constant at $T = 298$ K during the simulations, and the lateral pressure $P$ is increased stepwise at a fixed rate of 0.1 GPa/ns in the pressurization process. The sizes of graphene sheets in $z$ and $x$ directions are changed proportionally and chosen as 26.6 {\AA} $\times$ 21.9 {\AA}, 42.5 {\AA} $\times$ 35.0 {\AA}, 68.0 {\AA} $\times$ 56.0 {\AA}, 108.8 {\AA} $\times$ 89.6 {\AA}, respectively. From the local top views of final simulation snapshots, we can find that water molecules confined between graphene capillaries of different sizes all form the same square ice pattern, as shown in Figure \ref{fig:02}. The results of square ice structure are all in tune with the results in the previous study \cite{Algara2015SquareIce}, and further reveal that the phenomenon of square ice always happens, independent of the size of graphene within the range considered in this study. However, the corresponding critical crystallization pressure, stability of square structure, and time of crystallization process are significant affected by the different size of graphene sheets, which will be discussed in subsequent sections.

\subsection{Variation of critical crystallization pressure and unfreezable threshold}
The critical crystallization pressure, under which the square ice patterns begin to form, is one of the most important criteria to determine whether the phenomenon of square ice happens. Hysteresis simulation is shown in Figure S3 in Supporting Information, through a pressurization/depressurization process by increasing the lateral pressure until 2.0 GPa and then decreasing the lateral pressure back to 1 bar ($10^{-4}$ GPa). The pressurization/depressurization process exhibits a large hysteresis loop, implying that the formation of the square ice is a first-order phase transition. It was reported that, as the density increases, the nature of the phase transition inside the carbon nanotubes might change from a first-order transition to a continuous transition at the critical density \cite{han2010phase}. Overall, the pressurization/depressurization process triggers the freezing/melting transition, giving the compression/tension limit of the 2D water/ice, respectively. The critical crystallization pressure in this study is the compression limit of the 2D water.

\begin{figure*}[t]
  \centering
  \includegraphics[width=1.8\columnwidth]{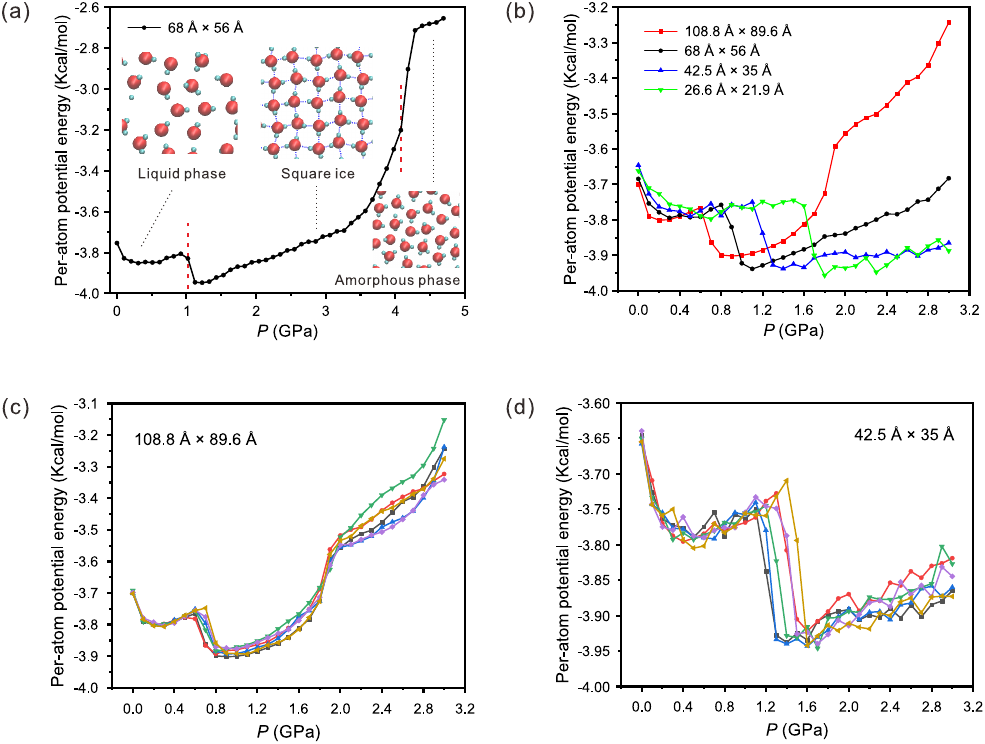}\\
  \caption{Variations of the per-atom potential energy for the confined water. (a) Phase transformation of water molecules from liquid to ice by increasing the pressure (the size of graphene sheets is 68.0 {\AA} $\times$ 56.0 {\AA}). The red globules indicate oxygen atoms, the cyan globules indicate hydrogen atoms, the blue dashed lines indicate hydrogen bonds and the red dashed lines are marks of the phase transformation. (b) Potential energy curves with different graphene sizes. (c) Variations of potential energy for repeated simulations with different initial velocities when the size of graphene sheets is 108.8 {\AA} $\times$ 89.6 {\AA}. (d) Variations of potential energy for repeated simulations with different initial velocities when the size of graphene sheets is 42.5 {\AA} $\times$ 35.0 {\AA}.}\label{fig:03}
\end{figure*}

To investigate the effect of graphene size on critical crystallization pressure, we performed simulations of pressurization process with graphene of different sizes. The lateral pressure $P$ was increased stepwise at a fixed rate of 0.1 GPa/ns in the pressurization process, and the variations of the potential energy for the confined water are shown in Figure \ref{fig:03} (Simulations with different pressurization rates are also performed and the results are shown in Figure S4 in Supporting Information, which show similar results). We can see that water confined in nanocapillary can transform from liquid to ice when the lateral pressure exceeds a critical value (i.e., approximately 1.0 GPa for graphene capillary of 68.0 {\AA} $\times$ 56.0 {\AA}, marked by red dashed line in Figure \ref{fig:03}(a)). In this study, all the results of critical crystallization pressure are obtained in three approaches: (1) The transition is accompanied by a sudden drop of the potential energy, so the potential energy curve can be used as an intuitive estimation of identifying phase transitions. (2) The square icing parameters, $MCV_1$ and $MCV_2$, are calculated to quantitatively distinguish between square ice structures and liquid water, and the same results of the critical crystallization pressure as the first method are obtained, as shown in Figure S5 in Supporting Information. (3) All the results of critical crystallization pressure are verified by observing the ice formation from simulation snapshots frame by frame, as shown in the snapshots of Figure \ref{fig:03}(a). However, the critical crystallization pressure of crystallization is different for graphene capillaries with different sizes, as shown in Figure \ref{fig:03}(b). Moreover, the critical crystallization pressure increases as the size of graphene capillary decreases, indicating that the size of the confinement significantly affects the formation condition of the square ice.

\begin{figure}
  \centering
  \includegraphics[width=\columnwidth]{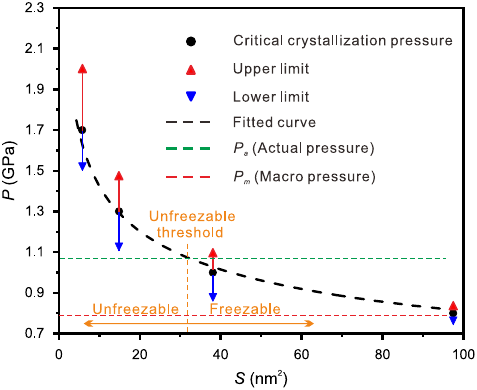}\\
  \caption{Critical crystallization pressure as a function of graphene size. The red arrows represent the upper limit of critical crystallization pressure, the blue arrows represent the lower limit of critical crystallization pressure, and the black dots represent the average critical crystallization pressure. The black dashed curve is the fitted curve of the average critical crystallization pressure, the green dashed line and red dashed line represent the actual pressure by estimating from the energy gain due to the squeezing effect and the macroscopic pressure (abbreviated as macro pressure), respectively. The orange dashed line perpendicular to the green and red dashed line represents the unfreezable threshold for the graphene size, below which water cannot freeze spontaneously.}\label{fig:04}
\end{figure}

To further analyze the effect of the graphene size and the variation of critical crystallization pressure of crystallization, we perform additional MD simulations. By changing the initial velocity distributions of the water molecules, the simulation for each graphene size is repeated six times. Figure \ref{fig:04} summarizes the critical crystallization pressure as a function of the graphene size ($S =$ $D_x$ $\times$ $D_z$), showing a strong dependence. As the graphene sheet becomes larger, the critical crystallization pressure slowly decreases, converging to a specific value ($\sim$ 0.8 GPa, red dashed line in Figure \ref{fig:04}). This situation corresponds to a quasi-macroscopic space (infinite in the $x$ and $z$ directions), thus we obtain the macroscopic crystallization pressure ($T$ = 298 K) from the phase diagrams of water (the phase diagram for the classical system described by the MLP, the ab initio phase diagram including NQEs, as well as the experimental phase diagram) for comparison \cite{reinhardt2021quantum}. The critical crystallization pressure is converging approximately to the macroscopic crystallization pressure ($\sim$ 0.6 -- 1.0 GPa) \cite{reinhardt2021quantum}. As the graphene sheet gets smaller, the critical crystallization pressure increases sharply, indicating that it is difficult for the confined water to form square ice pattern when the confining sheet is very small. It should be noted that the adhesion between the encapsulating graphene sheets imposes pressure on the water molecules as they are squeezed into a small volume. The actual pressure of water molecules $P_a$ can be estimated from the energy gain due to such squeezing effect ($P_a=E_w / d \approx 1.0$ GPa, where $E_w \approx 30$ MeV/{\AA}$^2$ is the difference in the adhesion energy per unit area between graphene-water and graphene-graphene, and $d \approx 5.6$ {\AA} is the increased separation distance between the two graphene sheets because of the trapped water molecules \cite{Algara2015SquareIce}). Therefore, there is an unfreezable threshold for the graphene size when the actual pressure is equal to the critical crystallization pressure. Even though this method is just a rough estimate and the value of $P_a$ is not precise, this result shows that there is an unfreezable threshold for the graphene size when the actual pressure is equal to the critical crystallization  pressure. It is hard to crystallize spontaneously in practice when the graphene sheet is smaller than the threshold, because the actual pressure is not high enough to match the critical crystallization pressure.

\subsection{Oscillation of critical crystallization pressure}
The results in Figures \ref{fig:03} and \ref{fig:04} also reveal that the critical crystallization pressure is not a fixed value with the same graphene size when the size is minuscule, but in a range of oscillation. The variations of potential energy for two of these graphene sizes are shown in Figures \ref{fig:03}(c) and (d), and results for other sizes are shown in Figure S6 and S7 in Supporting Information for details. When the size of graphene sheets is 108.8 {\AA} $\times$ 89.6 {\AA}, all curves of potential energy drop suddenly at almost the same pressure, which means that the critical crystallization pressure remains almost the same at this size of graphene sheets. But the critical crystallization pressure of crystallization changes significantly when the size of graphene sheets is minuscule, as shown in Figure \ref{fig:03}(d) for 42.5 {\AA} $\times$ 35.0 {\AA}. The critical crystallization pressure is in a range of oscillation from 1.1 to 1.6 GPa, which indicates that sometimes water molecules will crystallize under 1.1 GPa, and sometimes it will not until the pressure is up to 1.6 GPa. This phenomenon is similar to the range of temperature for the coexistence of nanoconfined ice and liquid \cite{Kastelowitz2018IceLiquidOscillations}. Figure \ref{fig:04} summarizes the oscillation of critical crystallization pressure as a function of graphene size. The red and blue arrows represent the upper and lower limits of the critical crystallization pressure, which are the maximum and minimum critical crystallization pressure from six repeated simulations (the distribution of critical crystallization pressure for all cases is shown in Figure S7). These upper and lower limits indicate the oscillation range of the critical crystallization pressure at a given graphene size. It shows a strong dependence of the oscillation range of the critical crystallization pressure on the size of the confining graphene. The oscillation range narrows as the sheets widen, converging to the macroscopic behavior of a single value of critical crystallization pressure for large sheets. We also summarize the critical crystallization pressure as a function of $D_z$, and the results show that there is no difference in the variation tendency between graphene size and critical crystallization pressure, as shown in Figure S8 in Supporting Information for details.

Finite-size effect is an important issue for molecular dynamics calculations of critical properties, and may have contributed to the variation and oscillation of critical crystallization pressure. In a finite system, the phase transition may be smeared over a pressure region, and the center of the pressure region may also be shifted, corresponding to the `rounding exponent' and the `shift exponent' \cite{binder1987finite}. In this study, the variations of the critical crystallization pressure correspond to the `shift exponent', and the oscillations of the critical crystallization pressure correspond to the `rounding exponent'. The `rounding exponent' and the `shift exponent' can be presented by many repeated simulation cases. In addition to the three methods used in this study, the critical crystallization pressure can also be obtained by some alternative methods, such as thermodynamic integration along a hypothetical path, which yields the relative free energies of the solid and liquid phases \cite{mastny2007melting}. In a previous study of the solid-liquid melting temperature, it has been shown that finite-size effects can account for deviations in the melting temperature (from the infinite-size limit) of up to 5$\%$ \cite{mastny2007melting}. The deviations in the critical crystallization pressure of square ice (from the infinite-size limit) are more than 112$\%$ in our study, which indicates that the finite-size effects account for only a small part of the pressure variation phenomenon.

\subsection{Stability of square icing patterns for all graphene sizes}

\begin{figure*}
  \centering
  \includegraphics[width=1.8\columnwidth]{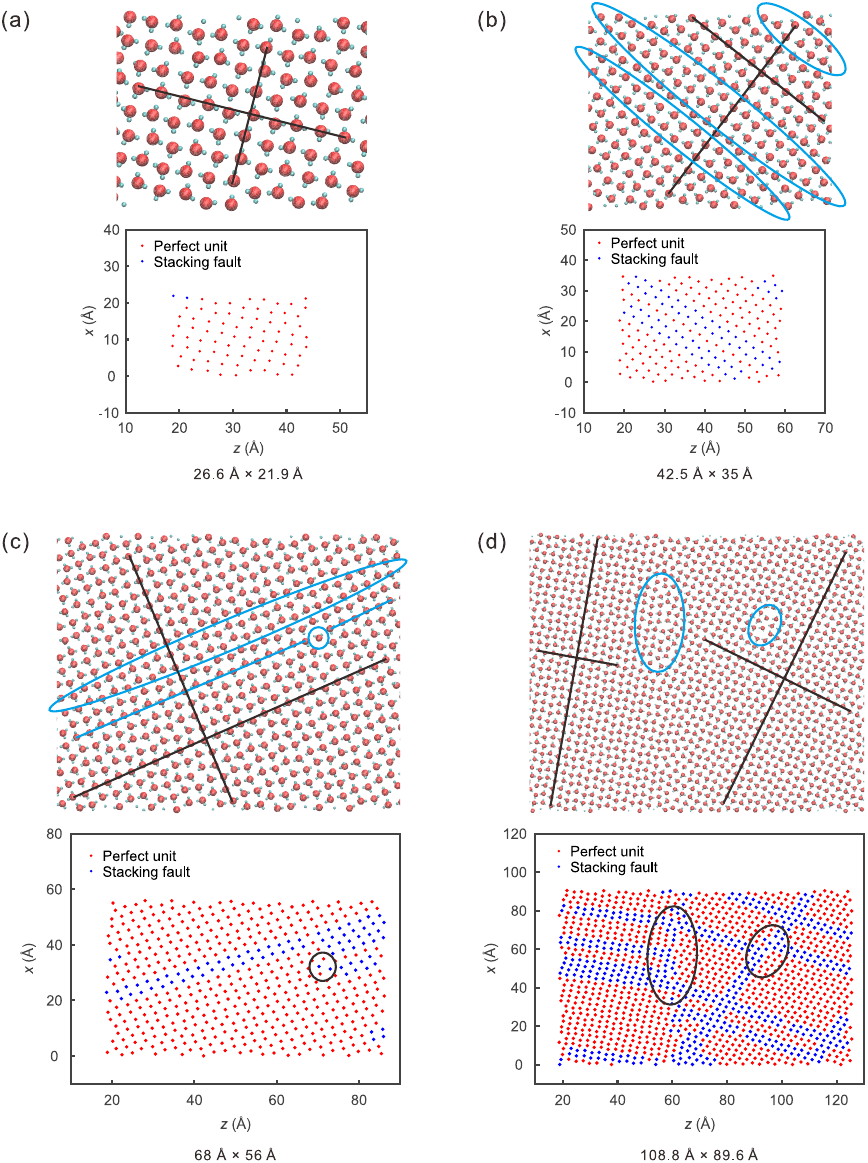}\\
  \caption{Top view snapshots of the simulation results and identification maps of crystal defects for graphene of different sizes. In the snapshots of the simulation results, the red globules indicate oxygen atoms, the cyan globules indicate hydrogen atoms, the black solid lines are marks of the icing direction and the blue graphics are marks of stacking faults and defects. In the identification maps of crystal defects, the red dots represent the ``perfect'' types of units and the blue dots represent the stacking faults and defects. The local defect is marked by the blue circles/ellipses in the top view and the black circles/ellipses in the map. (a) 26.6 {\AA} $\times$ 21.9 {\AA}. (b) 42.5 {\AA} $\times$ 35.0 {\AA}. (c) 68.0 {\AA} $\times$ 56.0 {\AA}. (d) 108.8 {\AA} $\times$ 89.6 {\AA}.}\label{fig:05}
\end{figure*}

Although water molecules confined between graphene of different sizes all form the same square icing pattern, the water molecules in our simulations still present differences in local structure. In square-like ice, the topological structure should satisfy the ice rule that one hydrogen atom must be donated for each water molecule along a column of the grid and the other along a row \cite{Koga2005HydrophobicSurfaces, Zhao2014FerroelectricHexagonal}. An elementary unit of square-like ice is made up of four water molecules and can be divided into four types \cite{Zhu2017MonolayerSquareIce}. In two of these types, which are the most common and ``perfect'', all water molecules in an elementary unit are of different arrangements. In the other two types, there are two or four water molecules with the same arrangement in an elementary unit (see Figure S9 in Supporting Information for a detailed classification of the elementary units). Square-like ice is mainly constituted by the first two ``perfect'' types of units. The regions constituted by the last two types of units can be regarded as stacking faults, which are a kind of defect in the crystalline structure. Therefore, a ``perfect'' square icing pattern in graphene nanocapillaries should be the alternatively distributed ``perfect'' units, without the same arrangement of water molecules in an elementary unit. However, many square ice structures are not perfect, and they can be regarded as assembled ``perfect'' units with stacking faults. Moreover, stacking faults can move, grow or dissipate in the crystallization process of square-like ices. In previous studies \cite{Bai2010MonolayerClathrate, Chen2016FirstPrinciples, Kaneko2013NewComputationalApproach, Kaneko2014HighDensityRhombic, Koga2005HydrophobicSurfaces, Mario2015AAStackedBilayer, Nair2012UnimpededPermeation, Qiu2013ElectromeltingMonolayer, Zhao2014TwoDimensionalAmorphousIce, Zhu2015CompressionLimit, Zhu2016AbStacked}, most MD simulation results satisfy this structural definition.

To show the local ice structure and the defects, the top view snapshots of the results and identification maps of crystal defects for graphene of different sizes are shown in Figure \ref{fig:05}. The stacking faults are marked by blue lines in the top view snapshots and by the blue dots in the identification maps of crystal defects. When the graphene size is relatively small, square icing patterns in nanocapillaries are composed of ``perfect'' units and stacking faults. As the size of graphene increases, the areas of stacking faults increase. Even so, all water molecules in graphene nanocapillaries still form an ordered structure for the relatively small size of graphene, as shown in Figures \ref{fig:05}(a), (b), and (c). As the graphene size increases further, the areas of stacking faults continue to increase, and water molecules in some area no longer form an ordered structure, as shown in Figure \ref{fig:05}(d) (i.e., the local defect marked by the blue circle in the top view and the black circle in the map). With the largest graphene sheets, there are many defects and areas with disordered states, which change the arrangement of protons nearby and divide the square icing pattern into several regions, as shown in Figure \ref{fig:05}(e) (i.e., the areas marked by the blue ellipses in the top view and the black ellipses in the map). The above results reveal that, as the size of graphene sheets increases, the stability of the square ice structure decreases, and the defects and disordered areas increase.

\begin{figure}
  \centering
  \includegraphics[scale=0.95]{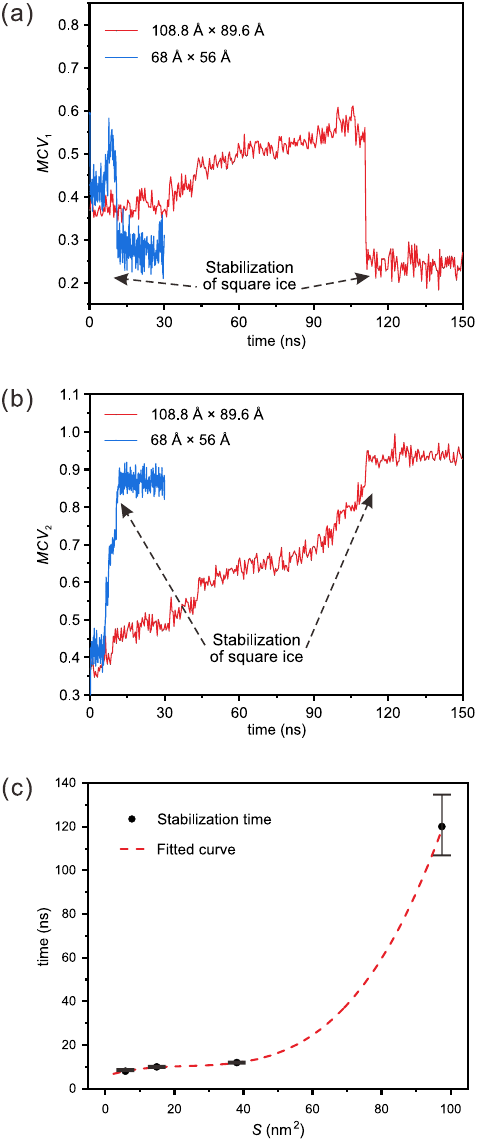}\\
  \caption{Comparison of crystallization times for all graphene sizes. (a) Comparison of crystallization times between two graphene sizes by the square icing parameters $MCV_1$. The red and blue curves represent different sizes of graphene sheets, 108.8 {\AA} $\times$ 89.6 {\AA} and 68.0 {\AA} $\times$ 56.0 {\AA}, respectively. (b) Comparison of crystallization times between two graphene sizes by the square icing parameters $MCV_2$. (c) Crystallization times for all graphene sizes. The black dots represent the average crystallization times and the red dashed curve is the fitted curve of the average crystallization time. Except for the size 108.8 {\AA} $\times$ 89.6 {\AA}, the errorbars for other sizes are too small to see.}\label{fig:06}
\end{figure}

For graphene sheets with different sizes, not only the stability of the square ice structure, but also the time for the square-like structure to stabilize is different. We can obtain the total time for the square ice to form until it is completely stable by analyzing the square icing parameters $MCV_1$ and $MCV_2$. These square icing parameters can quantitatively analyze the uniformity of the molecules distribution in the direction of the hydrogen-oxygen bonds and spatial position. Hence, they can be used to identify the different phases of water molecules. The comparison of the stabilization time for square ice formation between two graphene sizes is shown in Figures \ref{fig:06}(a) and (b). The stable values of these parameters at the beginning and the end of the curves indicate that water molecules are in two kinds of stable states, and the interval between them is the time of the stabilization process. From both $MCV_1$ and $MCV_2$, we can see obvious differences during the stabilization process for graphene of different sizes. With a larger graphene sheet, the stabilization process is slower, because of the larger crystallization area and the lower crystallization pressure. We calculate the stabilization time for all graphene sizes, as shown in Figure \ref{fig:06}(c). The stabilization time does not increase linearly with the graphene area, but increases sharply at large graphene size. It means that when the graphene sheet is large enough, it is very slow to form the square ice, even if the pressure required for crystallization is low.

\section{Conclusions}
In this study, we find that the critical crystallization pressure for the formation of square ice is not constant but strongly depends on the size of the graphene sheets, even at a fixed separation distance between the graphene sheets. The size of the graphene sheets is changeable over a wide range. Our results of all-atom molecular dynamics simulations reveal that the critical crystallization pressure of square ice slowly decreases as the graphene size increases, converging to approximately macroscopic crystallization pressure. As the graphene sheet gets smaller, the critical crystallization pressure increases sharply, indicating that it is difficult for the confined water to form the square icing pattern when the confining sheet is very small. There is a unfreezable threshold for graphene size when the actual pressure obtained by theoretical estimation is equal to the critical crystallization pressure, and it is difficult to crystallize spontaneously for square ice in practice when the graphene sheet is smaller than the threshold. Moreover, the critical crystallization pressure for the square ice formation is not a fixed value with the same graphene size but fluctuates when the graphene size is minuscule. The range of critical crystallization pressure oscillation also depends on the size of the confining graphene sheets, and it narrows as the sheets widen, converging to the macroscopic behavior of a single critical crystallization pressure for icing between large sheets. As the size of graphene sheets increases, the stability of the square ice structure decreases, the defects and disordered areas increase, and the stabilization time also increases. Hence, it will be very slow to form the square ice when graphene is large enough, even if the pressure required for crystallization is low. This study provides physical insights about the mechanism of critical crystallization pressure variation for the square ice formation, which is important for widespread applications in many fields such as nanomaterial, nanofluidic, and nanotribology.



\section{ASSOCIATED CONTENT}
\subsection{Supporting Information}
Further details of simulation results with different pressurization rates, variations of potential energy for all graphene sizes, classification of the elementary units, definitions and descriptions of the square icing parameters, including Equations and Figures (PDF).

\section{AUTHOR INFORMATION}
\subsection{Corresponding Authors}
Zhizhao Che - State Key Laboratory of Engines, Tianjin University, Tianjin 300072, China; Email: chezhizhao@tju.edu.cn\\
Tianyou Wang - State Key Laboratory of Engines, Tianjin University, Tianjin 300072, China; Email: wangtianyou@tju.edu.cn
\subsection{Authors}
Zhen Zeng - State Key Laboratory of Engines, Tianjin University, Tianjin 300072, China\\
Kai Sun - State Key Laboratory of Engines, Tianjin University, Tianjin 300072, China\\
Rui Chen - Department of Aeronautical and Automotive Engineering, Loughborough University, Loughborough LE11 3TU, United Kingdom\\
Mengshan Suo - State Key Laboratory of Engines, Tianjin University, Tianjin 300072, China

\subsection{Notes}
The authors declare no competing financial interest.

\begin{acknowledgement}
This work was financially supported by the National Natural Science Foundation of China (No.\ 51920105010 and 52176083).
\end{acknowledgement}


\small
\bibliography{Manuscript}



\end{document}